\documentstyle[12pt,epsfig]{article}
\def\haf{\textstyle{1\over2}}
\def\sevforth{\textstyle{7\over4}}
\def\bpi{\mbox{\boldmath{$\pi$}}}
\def\bsig{\mbox{\boldmath{$\sigma$}}}

\def\bt{{\bf t}}
\def\bq{{\bf q}}
\def\bA{{\bf A}}
\def\br{{\bf r}}
\def\bnabla{\mbox{\boldmath$\nabla$}}
\def\Nbar{{\bar{N}}}
\def\lsim{\mathrel{\raise3pt\hbox to 8pt{\raise -6pt\hbox{$\sim$}\hss{$<$}}}}
\def\rsim{\mathrel{\raise2pt\hbox to 9pt{\raise -7pt\hbox{$\sim$}\hss{$>$}}}}

\begin{document}
\vspace*{-1.25in}
\hspace{\fill} \fbox{\parbox[t]{2.0in}{\bf  LA-UR-97-4205  DOE/ER/41014-36-N97
THEF-NYM-97.02}}
\vspace*{0.5in}
\begin{center}

{\Large {\bf Electromagnetic Corrections to the One-Pion-Exchange Potential}}\\

\vspace{0.35in}
U.\ van Kolck$^a$, M.\ C.\ M.\ Rentmeester$^{b,e}$, J.\ L.\ Friar$^{c,d}$, \\
T.\ Goldman$^c$, and J.\ J.\ de Swart$^e$\\
\vspace*{0.20in}
$^a$ Department of Physics \\
University of Washington \\
Seattle, WA  98195-1560 \\
\vspace*{0.10in}
$^b$Department of Physics\\
The Flinders University of South Australia\\
Bedford Park, SA 5042, Australia\\
\vspace*{0.10in}
$^c$ Theoretical Division \\
Los Alamos National Laboratory \\
Los Alamos, NM  87545 \\
\vspace*{0.10in}
$^d$ Institute for Nuclear Theory\\
University of Washington\\
Seattle, WA 98195-1550\\
\vspace*{0.10in}
$^e$ Institute for Theoretical Physics\\
University of Nijmegen\\
Nijmegen, The Netherlands\\
\end{center}

\begin{abstract}
Leading-order electromagnetic loop corrections to the one-pion-exchange
potential are computed within the framework of chiral perturbation theory. These
corrections generate an effective nucleon-nucleon potential, $V_{\pi \gamma}$,
which supplements the sum of OPEP and the nucleon-nucleon Coulomb potential.
This potential is charge dependent and its construction is demonstrated to be
gauge invariant. The potential $V_{\pi \gamma}$ has been included in the
Nijmegen partial-wave analysis of $np$ data. A particular renormalization scheme
is chosen that leads to a negligible change in the $\pi^{\pm}N\!N$ coupling
constant and in the $np \; ^1\!S_0$ scattering length and effective range.
\end{abstract}

\pagebreak

Isospin violation is an august topic\cite{1,2} in nuclear physics, and
considerable progress has been made in understanding the consequences of isospin
violation in nuclei, particularly in few-nucleon systems\cite{3,4}.
Nevertheless, a comparable understanding of the underlying mechanisms for
isospin violation in the nuclear force has been slower in developing.  It is our
purpose here to present for the first time complete analytic results for one
such mechanism obtained using chiral perturbation theory, and to determine its
effect on the $\pi^\pm N\!N$ coupling constant and the $np \; ^1\!S_0$
scattering length and effective range.

It has been known for decades that the various mechanisms responsible for
isospin violation could be graded in strength, with charge-dependent (CD)
nuclear forces being generally stronger than charge-symmetry-breaking (CSB)
nuclear forces, while strongest of all is the long-range Coulomb force between
protons. Only recently, however, have these empirical observations been linked
quantitatively to microscopic strong-interaction mechanisms based on 
symmetries\cite{5,6,7}.

The strength of isospin violation can be understood\cite{5} using dimensional 
power counting, a property associated with chiral effective Lagrangians\cite{6}.
This technique expresses the strength of any nuclear interaction in terms of 
several energy scales: a heavy scale $M_{\rm QCD}\sim$ 1 GeV (the 
characteristic QCD mass) and the lighter scales $f_{\pi} =$ 92.4 MeV (the pion 
decay constant), $m_{\pi} = 139.6$ MeV (the scale of explicit chiral-symmetry 
breaking), and $Q$ (the effective momentum in a nucleus, which can be taken to 
be $\sim m_{\pi}$). In addition, up-down quark-mass-difference-induced isospin 
violation\cite{7} carries an extra factor of $\epsilon = (m_d - m_u)/(m_d + 
m_u) \sim$ 0.3, while electromagnetic interactions carry powers of the 
fine-structure constant $\alpha$ ($\sim$ 1/137). Dimensionless 
strong-interaction constants of ``natural''\cite{8} size $\sim$1 (such as $g_A 
= $ 1.26, the axial-vector coupling constant) are ignored. The general 
technique is not dependent on any particular model, but specific force models 
can be analyzed and expressed in terms of these quantities, as well. Signs are 
not determined by such arguments.

Using these techniques one finds that the dominant (isospin-conserving)
component of the nuclear force in light nuclei, OPEP, has a strength\cite{9}
(corresponding to an energy shift) $\sim$ 15 MeV/nucleon pair, while the $pp$
Coulomb interaction (CD and CSB) is roughly 1 MeV in strength.  The small Breit
corrections to the latter (viz., magnetic and velocity-dependent forces) have a
nominal 20 keV strength, as does the effect of the nucleon-mass difference.  The
difference in pion masses in OPEP generates a CD nuclear force of roughly 1/2
MeV in strength.  Simultaneous exchange of a pion and a photon between nucleons
has a nominal 30 keV strength, as does the electromagnetic modification (at the
quark level)\cite{10} of the $\pi N\!N$ coupling constant; both are CD
mechanisms.  The leading-order CSB force generated by the effect of differing
quark masses on the $\pi N\!N$ coupling constants may be somewhat larger $\sim$
90 keV. These estimates (which could easily vary by a factor of two) correspond
rather well to those obtained from the well-studied isospin violation in
few-nucleon systems\cite{3,4,10}. Thus, one finds that CD is larger than CSB,
and both of these mechanisms are larger than the tiny isospin-violating forces
of Type IV \cite{1}, all of which was first demonstrated in Ref.\ \cite{5} using
dimensional power counting.

Isospin violation in the nuclear force has also been studied as a natural
extension of the partial-wave analysis (PWA) of both $pp$ and $np$ scattering
data\cite{11}, together with the tiny amount of $nn$ scattering
information\cite{12}.  It has long been known that the $NN \; ^1\!S_0$
scattering lengths are all different, and their differences scale roughly as the
interaction strengths estimated above.  These PWAs have allowed the $\pi^0 pp$,
$\pi^0 np$, and $\pi^\pm N\!N$ coupling constants to be separately
determined\cite{11x}, and the Goldberger-Treiman (GT) discrepancy\cite{13} to be
obtained, as well.  At the present time\cite{11x,10} both the CD and CSB $\pi
N\!N$ coupling constants are consistent with zero (within $\lsim$ 1\% overall
uncertainty in the $\pi N\!N$ constants), and the GT discrepancy $(d-1)$, which
is isospin conserving but chiral-symmetry breaking, is approximately 2\%.

The Nijmegen PWA\cite{11} incorporates pion-mass differences in OPEP ($V_\pi$),
nucleon-mass differences, and the full electromagnetic interaction\cite{13x}
(including vacuum polarization) when it determines the $\pi N\!N$ coupling
constants. Only the effective $\pi$-$\gamma$-exchange interaction has been
missing, and we have seen that its strength is comparable to that of several
isospin-violating mechanisms normally incorporated. The nominal strength of that
force is (as we shall see) $(\alpha/\pi) V_{\pi}$, and this is comparable to the
uncertainty in the $\pi^\pm N\!N$ coupling constant, $f^2_c = (g_A\, m_{\pi^+}
\, d / 2 f_\pi )^2 / 4 \pi$.  Therefore, the task that remains is to calculate 
the effective $\pi$-$\gamma$-exchange force, incorporate it into the Nijmegen 
PWA, and redetermine $f^2_c$.

The calculation is performed within the framework of chiral perturbation
theory\cite{5,6,14} to leading order in $1 / M_{\rm QCD}$.  This allows us to
make the static approximation for nucleons $(M_N \rightarrow \infty$), and
restricts us to the leading-order terms (in a chiral expansion) in the effective
Lagrangian.  Because we are generating EM corrections to one-pion exchange
between nucleons (short-range forces were discussed previously\cite{5,10}), only
single-loop corrections involving photons are required.  In static order, the
necessary elements of the leading-order interaction Lagrangian are:
\begin{eqnarray*}
\hspace*{0.5in} L^{(0)} &=& 
-\frac{g_A}{f_\pi} \, \Nbar [ \bsig \cdot \bnabla
(\bt \cdot \bpi) ] N +\frac{e g_A}{f_\pi}  \Nbar \bsig \cdot \bA
\, (\bt \times \bpi)_3 N \\
&{}& - e A_0 \Nbar (\haf+t_3) N \, - e A^{\mu} (\bpi \times \partial_{\mu} 
\bpi)_3 + \haf e^2 A^2 (\bpi^2 -\pi_3^2) + \cdots \, , \hspace*{0.23in} (1)
\end{eqnarray*}
where we have used Cartesian notation for the pion charge states ($\pi_i$) but
have otherwise conformed to the notation and conventions of Refs.\ \cite{15} and
\cite{10}. Additional terms in the Lagrangian that are not required here can be
found in the latter reference. The nucleon (Pauli) spin operator is $\bsig$, its
isospin operator is $t_{\alpha} (t_3 = \pm \haf)$, $e$ is the fundamental
(proton) charge, and $A^{\mu}$ is the photon field.  Although to leading
(static) order the $\pi \gamma N$ seagull interaction involves only $\bA$ and
the nucleon EM interaction involves only $A_0$, the full pion EM interaction is
required.

\begin{figure}[ht]
  \epsfig{file=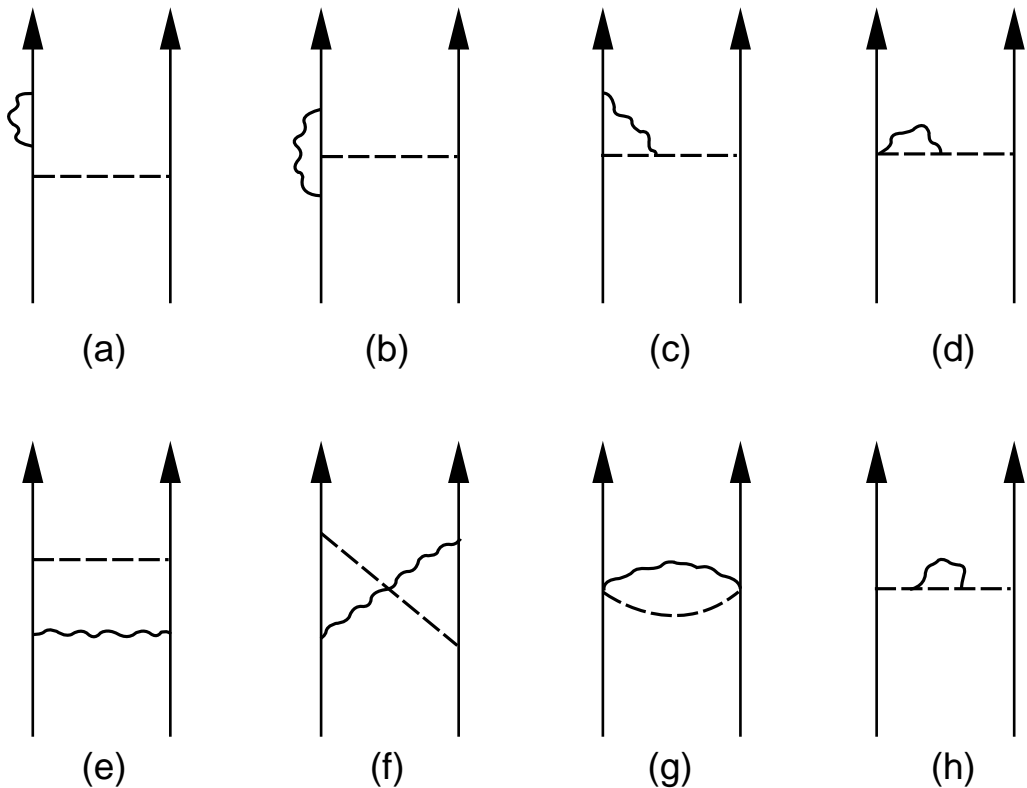,height=4.5in}
  \caption{Electromagnetic loops contributing to isospin violation in the
           OPE nuclear force in leading order. Solid lines are nucleons, 
           dashed lines are pions, and wavy lines are photons.}
\end{figure}

The calculation requires all diagrams of Fig.\ (1).  Figure (1a) subsumes 4
separate diagrams (a bubble on each nucleon leg), (1b) 2 diagrams, (1c) 4
diagrams, (1d) 2 diagrams, (1e) and (1f) each 2 diagrams, (1g) is unique, while
(1h) is accompanied by an EM tadpole (obtained from the last term in Eq.\ (1))
that is not shown.  Altogether, 19 diagrams were separately calculated in both
Coulomb and Feynman gauges.  All graphs differ in the two gauges.  Coulomb gauge
has the advantage of eliminating {\it a priori} any infrared divergences, but
considerable complexity is introduced into the calculation.  Infrared
divergences arise in Feynman gauge but cancel, while a complicated set of terms
that ultimately cancel in Coulomb gauge never arise in Feynman gauge.  The sum
of diagrams is the same in both gauges, a natural consequence of any complete
calculation, rather than one focusing on a few selected mechanisms.

In order to obtain physical and useful results we have to use a renormalization
prescription.  We follow Ref.\ \cite{16} and keep only nonanalytic terms in
divergent loops, incorporating all analytic terms into the definitions of the
coupling constants and masses. The hard parts of vertex loops renormalize the CD
$\pi N\!N$ coupling constant (denoted by $\bar{\beta}_{10}$ in Refs.\
\cite{5,10}), for example. In addition, diagrams (1e) and (1f) contain a
component corresponding to the iteration of the static OPEP and Coulomb
potential (in the general case) and this has been subtracted from the full
amplitude, since it is automatically included when solving the Schr\"{o}dinger
equation.  The remaining amplitude $(S_{\pi \gamma})$ is still gauge invariant
and the relation
$$
S_{\pi \gamma} = - i\, V_{\pi \gamma}  \eqno (2)
$$
defines the effective $\pi$-$\gamma$-exchange potential, $V_{\pi \gamma}$,
between two nucleons arbitrarily labelled 1 and 2. Its isospin structure allows
only charged-pion exchange and therefore affects only $np$ scattering. The usual
OPEP corresponding to charged-pion ($\pi_c$) exchange, $V_{\pi_c}$, is similarly
obtained.  Their sum is given by the simple momentum-space expressions
$$
V_{\pi \gamma}(\bq) + V_{\pi_c}(\bq) = - \frac{g_{A}^{2}} { f_{\pi}^{2} \, 
   m_{\pi}^2}(\bt_1 \cdot \bt_2 - t_1^3 \, t_2^3)  (\bsig_1 \cdot \bq \, 
   \bsig_2 \cdot \bq) \, [V_{\pi \gamma}(\beta) + V_{\pi_c}    (\beta)] \, , 
   \eqno (3)
$$
$$
V_{\pi \gamma}(\beta) = \frac{\alpha}{\pi} \left[- \frac{(1- \beta^2)^2}
{2 \beta^4 (1+ \beta^2)} \ln (1 + \beta^2) +\frac{1}{2 \beta^2} - 
\frac{2\, \bar{\gamma}}{1+\beta^2 } \right] \, , \eqno (4)
$$
where OPEP is determined from
$$
V_{\pi_c}(\beta) = \frac{1}{1 + \beta^2} \, , \eqno (5)
$$
and $\beta = q/m_{\pi}, \bq$ is the momentum transferred between the nucleons,
$m_{\pi}$ is the charged-pion mass, and $\bar{\gamma}$ will be discussed below. 
Details of this calculation (including graph-by-graph results), pion and nucleon
$\sigma$-terms, and the renormalization-scale dependence of the coupling
constants will be published elsewhere.

This force is CD and remarkably simple in form, given the number of processes
that contribute.  Fourier transformation to configuration space yields
$$
V_{\pi \gamma}(\br)  = \frac{g_{A}^{2} \, m_{\pi}^{3}}
   {4\pi f_{\pi}^{2}} \left[ \frac{\alpha}{\pi} \right]
   (\bt_1 \cdot \bt_2 - t_1^3 \, t_2^3) \,
    \bsig_1 \cdot \bnabla_{\! z} \, 
    \bsig_2 \cdot \bnabla_{\! z} \, 
    \left[ \frac{I (z)}{z} \right] \, , \eqno (6)
$$
$$
I (z)  =  2 e^{-z} [\, \ln (z/2) + \gamma_E -\bar{\gamma}] + 2e^{z} Ei (-2 z)
 - Ei (-z) (3 + \frac{z^2}{2}) + \frac{e^{-z}}{2} (1-z) \, ,
\eqno (7)
$$
where $Ei(-z) = - \int^{\infty}_z d t \, e^{-t}/t$ is the exponential integral,
$z = m_{\pi}\, c\, r / \hbar$, $\gamma_E$ is Euler's constant (0.577$\cdots$), 
and the usual OPEP corresponding to charged-pion exchange is obtained
by substituting $e^{-z}$ for $I(z)$ and dropping the factor of $\alpha / \pi$ in
brackets.  Our renormalization prescription defines $\bar{\gamma}$ and is
discussed next.  The first term in $I(z)$ (in brackets) determines the
asymptotic form, while the volume integral of $I(z)$ is proportional to
($\sevforth- 2 \bar{\gamma}$).  A positive value of $\bar{\gamma} \, (\sim 1)$
weakens both.

The usual definition\cite{11x} of the $\pi N\!N$ coupling constant requires an
extrapolation to the pion pole (i.e., $q^2 = m^2_{\pi}$) in the unphysical
region of the NN scattering amplitude, with the residue defining the coupling
constant and the difference between $q^2 = 0$ and $q^2 = m^2_{\pi}$ defining the
GT discrepancy\cite{13}.  This pole corresponds (since $q^0 = 0$ in the static
limit) to $\beta^2 = -1$.  Although $V_{\pi_c} ( \beta )$ has a simple pole at
$\beta^2 = -1$, the first term in $V_{\pi \gamma} ( \beta )$ does not (the
residue diverges) because of the logarithm induced by the infrared structure of
the photon loops.  This is reflected in the configuration space term:  $e^{-z}
\ln (z)$.  Any constant multiple of $e^{-z}$ in $I(z)$ (i.e., a multiple of
OPEP) can be arbitrarily transferred to OPEP with an appropriately redefined
$f^2_c$, since the sum of $V_{\pi \gamma}$ and $V_{\pi_c}$ remains unchanged.
Alternatively, the logarithmic (asymptotic) term in $V_{\pi \gamma} (\beta)$ can
be made to vanish at any convenient point.  Straightforward development of
$V_{\pi \gamma} (\beta)$ leads to $\bar{\gamma} = 0$ (an even simpler result!). 
We choose, however, to remove the $\gamma_E$ term in $V_{\pi \gamma}(\br)$ by
performing a further finite renormalization and fixing $\bar{\gamma} \equiv
\gamma_E$; this {\bf defines} $f_c^2$ in the presence of EM corrections, and is
analogous to the $\overline{MS}$ (Modified Minimal Subtraction) renormalization
commonly used in Standard Model calculations. The bracketed (logarithmic) term
in $I ( z )$ now vanishes at $r = 2 \hbar / m_{\pi} c = 2.8$ fm.  As we have 
previously discussed, this weakens both the tail of the $\pi$-$\gamma$ potential
and the volume integral of $I(z)$ (by a factor of three).

It is difficult to compare our result with previous calculations\cite{17,18,19}.
None of the CD calculations\cite{20x,20,21,22,23,24} were complete, and few were
written in an easily interpretable form.  Some numerical results are available,
however\cite{25}. The gauge invariance of our final result gives us confidence
in its form. Figure (1g) is easily calculated in old-fashioned perturbation
theory in the static limit (see Ref.\ \cite{25}), and a comparison checks our
overall sign and factors.  As an additional check, the Breit
interaction\cite{26} can be calculated using our conventions and yields the
usual result.

\begin{table}[htb]
\centering
\caption{Nijmegen PWA fits. The $\pi^\pm N\!N$ coupling constant, $f_c^2 (\times
1000)$, and the corresponding $\chi^2$ of fit for a variety of cases are
indicated in the top row. The potentials are those included in the tail of the
$np$ force, while the quantities in brackets were those fit to produce the
results listed directly below. ``All'' denotes that the parameterized interior
region of the force was fit in addition to $f_c^2$.}

\hspace{0.25in}

\begin{tabular}{|l||ccc|}
\hline
{Fit}\rule{0in}{2.5ex} &$V_{\pi}\; [\rm{all}]$&$V_{\pi} + V_{\pi \gamma} \; 
[f_c^2]$ & $V_{\pi} + V_{\pi \gamma}\; [\rm{all}]$ \\ \hline \hline
$10^3 f_c^2$ \rule{0in}{2.5ex} & 74.96(34) & 75.22 & 74.98(33)\\ 
$\chi^2$ & 4223.6 & 4236.5 & 4222.8 \\ \hline
\end{tabular}
\end{table}

The tail of $V_{\pi \gamma}$ (for $r >$ 1.4 fm) was incorporated into the
Nijmegen PWA of only the $np$ data\cite{27} on an equal footing with OPEP.  A 
total of 4107 $np$ data were fit, and the results are shown in Table I.  Each
entry lists the $\chi^2$ of the fit and the fitted value of $f^2_c (\times
1000)$. The first entry is the 1997 Nijmegen $np$ PWA result~\cite{27}. Simply
adding $V_{\pi \gamma}$ increases $\chi^2$ by a factor of 8. This can be greatly
reduced by refitting only $f^2_c$, which is shown in the interim (next) entry.
Refitting both $f^2_c$ and the phenomenological interior region ($r <$ 1.4 fm)
leads to the rightmost entry. Only those entries labelled ``all'' should be
compared.

Although it is necessary to increase slightly the strength of the tail of OPEP
to compensate for the addition of $V_{\pi \gamma}$ in the overall fit, there is
negligible change in $f^2_c$, and only a tiny decrease in $\chi^2$.  The $np \;
^1\!S_0$ scattering length and effective range also show negligible change.  We
conclude that the addition of $V_{\pi \gamma}$ has not affected any of the
former conclusions of the Nijmegen PWA with regards to $f^2_c$, quality of fit,
or low-energy scattering observables. We emphasize that this does not mean that
$V_{\pi \gamma}$ is everywhere negligible, since that part with $r < 1.4$ fm is
subsumed in the phenomenological interior region of the Nijmegen PWA. If a force
were to be constructed (for all $r$) with an explicit $\pi \gamma$ component,
the effect of the latter would be considerably larger. This $np$ force is
repulsive in S-waves for $r < 3.7$ fm, but attractive otherwise. A rough 
estimate using {\it ad hoc} form factors and several $pp$ potentials shows that 
the {\it magnitude} of the $np \; ^1\!S_0$ scattering length decreases by 
$\sim$ 0.67 fm. This is twice the strength of the effect found earlier from 
the double-seagull mechanism\cite{24}. The deuteron energy is raised by roughly
60 keV.

As a check of our procedures, we have also fitted the cases $\bar{\gamma} =
\gamma_E \pm 1$. These alternatives to our renormalization scheme either add or
remove a specified fraction ($2 \alpha/\pi$) of OPEP from $V_{\pi \gamma}$, and
this must be compensated in the final fit by a corresponding change in $f^2_c$.
This is indeed found.  Choosing $\bar{\gamma} = 0$ (no finite renormalization)
lowers $1000 f^2_c$ by $0.18$, which is slightly more than half the quoted
uncertainty.  We elect, however, to use our preferred convention ($\bar{\gamma}
= \gamma_E$), since that leaves $f^2_c$ essentially unchanged.

In summary, we have calculated 19 Feynman graphs that produce the leading-order
one-loop EM corrections to OPEP.  The results are gauge invariant and remarkably
simple, and generate a static charge-dependent $\pi$-$\gamma$-exchange
potential.  This potential, which has a nominal strength $(\alpha / \pi) V_\pi$,
was one of the few isospin-violating mechanisms not previously incorporated
into the tail of the NN force by the Nijmegen PWA.  When incorporated using our
renormalization prescription, there is negligible change in the $^1\!S_0$
low-energy parameters and the $\pi^\pm N\!N$ coupling constant, $f^2_c$, and
only a tiny improvement in the quality of fit.

{\Large {\bf Acknowledgments}}

The work of UvK, JLF and TG was performed under the auspices of the United 
States Department of Energy. MR acknowledges the financial support of the 
Australian Research Council. We would like to thank S.\ -N.\ Yang for several 
useful conversations. One of us (UvK) would like to thank G.\ A.\ Miller, R.\ 
B.\ Wiringa, and E.\ M.\ Henley for several useful discussions, while JLF 
acknowledges the help of S.\ A.\ Coon.

\end{document}